\documentclass[10pt]{iopart}
\usepackage{iopams}
\usepackage{graphicx}% Include figure files
\usepackage{dcolumn}% Align table columns on decimal point
\usepackage{bm}% bold math
\usepackage{color}
\usepackage{amsbsy}
\usepackage{amssymb}
\usepackage{subfig}
\usepackage{wasysym}
\usepackage{booktabs}

\newcommand{\nn}{\nonumber}
\newcommand{\be}{\begin{equation}}
\newcommand{\ee}{\end{equation}}
\newcommand{\ba}{\begin{eqnarray}}
\newcommand{\ea}{\end{eqnarray}}

\def\n{\vskip.2cm \noindent}

\begin{document}

\bibliographystyle{prsty}

\title[]{Optimizing the Earth-LISA ``rendez-vous''}

\author{Fabrizio De Marchi$^1$\footnote[7]{Corresponding author: fdemarchi@science.unitn.it}, Giuseppe Pucacco$^{2}$, Massimo Bassan$^{2}$}

%, Stefano Vitale$^1$

\address{$^1$ Department of Physics, Universit\`a di Trento and INFN, Sezione di Trento, I-38100 Povo}

\address{$^2$ Dipartimento di Fisica, Universit\`a di Roma ``Tor Vergata'' and INFN, Sezione di Roma Tor Vergata, I-00133 Roma}

\begin{abstract}
We present a general survey of heliocentric
LISA orbits, hoping it might help in the exercise of
rescoping the mission. We try to semi-analytically optimize the orbital parameters in order to minimize the disturbances coming from the Earth-LISA interaction. In a set of numerical simulations we include non-autonomous perturbations and provide an estimate of Doppler shift and breathing as a function of the trailing angle.
\end{abstract}


\section{Introduction}\label{intro}
The LISA space experiment to detect low frequency gravitational waves has been for a long time a priority mission of space agencies, both in Europe and in the US.
There has recently been an ample discussion on a possible scaled-down version of the LISA mission that, in order to meet tighter budget constraints, could be characterized by a shorter arm length $L$, a closer mean distance from the Earth (a smaller trailing angle) and maybe a 2-arms (4-links) configuration, giving up the third arm. In this case, it becomes natural to consider a right angle geometry as an alternative to the traditional, $60^\circ$, equilateral triangle.

\n
{Although other configurations are being evaluated by the ESA Concurrent Design Facility team, these triangular `constellations' on heliocentric Earth-trailing orbits still remain the favorite choice. We focus our attention on the evaluation of the usual kinematic indicators of performance (arm flexing, breathing angles and Doppler shifts) when reducing both the size of the triangle and the Earth-LISA distance over the entire mission lifetime. As is well known, the interaction of LISA with the Earth is the major perturbation. The dominant effect is a parabolic drift characterized by a ``rendez-vous'' (RV) at which the distance between the constellation and the Earth is minimum. We investigate how additional perturbing effects influence the motion of LISA around the RV and how it is possible to optimize it.}

\n
We assume the following guidelines:
\begin{itemize}
\item Arm Length:  $L$=1 Gm. We consider two configurations: the equilateral triangle (ET) with side $L$ and an isosceles right triangle (IRT) with two equal arms of length $L$ and the third one $L \sqrt{2}$ long. 
\item  Flexing:  the spacecraft (S/C)  relative velocity in the sensitive axis  (rate of change of the arm length) causes a Doppler shift of the laser frequency. We set a maximum bandwidth of 20 MHz over a 1$\mu$m carrier, corresponding to $V/c < 6.5 \times 10^{-8}$, or $V < 20$ m/s.  Note, for the sake of plot readability, that the Doppler shift (in MHz) has the same numerical value than the longitudinal velocity (in m/s).
\item Breathing angle: The relative motion of S/C's also imposes a continuous adjustment of the angle between two beams departing from the same corner, in order to track the opposite spacecrafts; this fluctuation over the nominal angle ($60^{\circ}$ or $90^{\circ}$) is referred to as breathing angle (BA). We demand BA $< \pm 1.5^{\circ}$.
\item  Trailing angle (TA, also referred to as Lag Angle): {LISA follows the Earth on {a heliocentric circular orbit} and TA is the angle between the constellation and the Earth as seen from the Sun.} {TA is a good indicator of the Earth-LISA distance, because the radial secular motion (away from the Sun) of the constellation is normally much smaller than the tangential (along the Earth orbit) one.}  We demand TA as small as possible, compatibly with the above requirements. {In any case, initial conditions are chosen in such a way that over the mission lifetime, TA never exceeds 21$^{\circ}$.} 
\item  Mission lifetime:  6  years.
\end{itemize}

\n
The plan of the paper is as follows: we start recalling the simple models describing the interaction between Sun and LISA (\Sref{kepl}) and Sun, Earth and LISA (\Sref{EE}). In \Sref{MIN} we describe an optimization method with the aim of an important reduction of flexing, breathing angles and Doppler shifts. We test this method first on the simplified Sun-Earth-LISA model and then on a more complete model including the real gravitational effects due to the dynamics of the Solar System. Finally, in \Sref{conclusions} conclusions are drawn.

\n
While graphs and details are given below, we anticipate here some
results:
\begin{itemize}
\item As far as Doppler and breathing requirements are concerned, a short LISA can safely be put in an orbit much closer to Earth: TA $ \approx12^{\circ}$ {at RV (the
baseline design was  $20^{\circ}$), or 31 Gm}.  
\item Should we give up the third arm (keep only 4 optical links), a right angled
triangle can be employed and performs at least as well as the usual equilateral
triangle in several of the tests (Doppler, breathing, etc.) we carried on. \end{itemize}

\section{Keplerian orbits}\label{kepl}
We describe the interaction of LISA with the Earth in the framework of the Hill-Clohessy-Wiltshire (HCW) system \cite{bvj,nerem}. {In this and next section, we will assume the Sun at rest in an inertial reference frame}. The  origin of the HCW frame rotates around the Sun on a circular reference orbit of radius $R_0$ and with orthogonal axes oriented as follows: $x$ is directed radially opposite the Sun, $y$ is in the direction tangent to the motion and $z$ is perpendicular to the ecliptic. The time evolution of these orbits can be described with adequate precision using the post-epicyclic approximation  in the HCW frame \cite{cqg2,sw}.

\subsection{Zero order approximation}
Under the effect of the Sun only, at zeroth order, the equations of motion for a S/C in the rotating frame are

\begin{equation}
\eqalign{
 \ddot x -2\omega \dot y -3\omega^2 x = 0,\cr
 \ddot y +2\omega \dot x =0,\cr
 \ddot z+\omega^2 z = 0,
}\label{eqhcw}
\end{equation}
where $\omega = \sqrt{G M_{\odot} / R_0^3}$ is the mean motion and the most general solution is a combination of an ellipse in the $xy$ plane and an oscillation in the $z$ direction

\begin{equation}
\eqalign{
\hspace{-2cm}x (t)  =  2\left(2 x_0+ \frac{\dot y_0}{\omega}\right)-\left(2 \frac{\dot y_0}{\omega}+3 x_0\right)\cos \omega t +\frac{\dot x_0}{\omega} \sin \omega t, \cr
\hspace{-2cm}y ( t) =  y_0-2\ \frac{\dot x_0}{\omega}-3 \,\left(\dot y_0+2 \omega x_0 \right) t\, +2\frac{\dot x_0}{\omega} \cos \omega t +2 \left(2 \frac{\dot y_0}{\omega}+3 x_0\right)\sin \omega t, \cr
\hspace{-2cm}z (t)  =  z_0 \cos  \omega t  +\frac{\dot z_0}{\omega} \sin  \omega t .
}
\label{soleqhcw}
\end{equation}
The natural choice, in order to avoid drifts and offsets, is to set $\dot y_0=-2 \omega x_0$ and $\dot x_0=\omega y_0/2$, so that 
 the trajectory is reduced to a combination of simple oscillations along the three axes.
Moreover, for a rigid, polygonal constellation, the distance of the S/C from the origin must be constant, say $h$, so that we obtain \cite{cqg1}

\[
\hspace{-2.25cm} 
\dot z_{0}= {\pm} \frac{\sqrt{3}}{2} \omega y_{0}; \quad z_{0}={\pm} \sqrt{3} x_{0}; \quad y_0={\pm} \sqrt{h^2-4 x_0^2};  \quad x_0=\pm \frac{h}{2}.
\]

\n
The orbit of one of the S/C's in the HCW frame is a circular motion with constant angular velocity $\omega$ and radius $h$ around the origin in a plane inclined of $\pm 60^\circ$ with respect to the $xy$ (ecliptic) plane.
A second S/C, describing the same path (with a certain delay), will be at a constant distance  $\ell$, from the first one. For $n$  such S/Cs, on the vertices of a regular polygon,  their distance $h$ from the origin and the relative phase delay are

\[
h=\frac{\ell}{2 \sin (\pi/n)},\quad  \phi=\frac{2\pi}{n}.
\]
Finally, we put $x_0=\ell/2$ to be consistent with the notations of \cite{auto,cqg2} and the zero-order orbits turn out to be

\be\label{zerovec} \label{rzero}
\hspace{-2cm} \mathbf{r}^{(0)}_{k}(t)=\frac{\ell}{2 \sin (\pi / n)} \left[ \frac{1}{2} \cos \sigma_k,\sin \sigma_k,\frac{\sqrt{3}}{2}\cos{\sigma_k}\right],\quad \sigma_k=\frac{2\pi (k-1)}{n}-\omega t.
\ee
The ET constellation is obtained with $n=3$ and $k=1,2,3$, while the IRT one corresponds to $n=4$ (and $k=1,2,3$), namely

\begin{equation*} 
\eqalign{
\hspace{-2cm} \mathbf{r}^{(0)}_{k}(t)=\frac{\ell}{\sqrt{2}} \left[ \frac{1}{2} \cos \sigma_k,\sin \sigma_k,\frac{\sqrt{3}}{2}\cos{\sigma_k}\right],\quad \sigma_k= (k-1)\frac{\pi}{2}-\omega t.
}
\end{equation*}

\subsection{First order approximation}\label{firstord}
 A first optimization is possible by changing the tilt angle of the constellation, i.e. the inclination of the triangle with respect to the ecliptic. Introducing a parameter $\delta_1$ \cite{cqg2} such that:

\[
\pm 60^\circ + \delta_1 \frac{\ell}{2R_0}.
\]
the first order corrections 
\begin{equation}\label{runo}
\mathbf{r}^{(1)}_{k}(t)=(x^{(1)}_{k},y^{(1)}_{k},z^{(1)}_{k}),
\end{equation}
are:
\begin{equation}
\eqalign{
x^{(1)}_{k}(t)= \frac{h^2}{2R_0}\left[ \frac{3}{2}\left(\frac{1}{2}-\delta _1\right) \cos \sigma _k-\frac{1}{8} \cos \sigma
   _k-\frac{5}{8} \right],\cr
 y^{(1)}_{k}(t)= \frac{h^2}{2R_0}\left[ \left(\frac{3}{2}-3 \delta _1\right) \sin \sigma _k-\frac{1}{2} \sin 2
   \sigma _k \right],\cr
z^{(1)}_{k}(t)= \frac{h^2}{2R_0}\left[\frac{\sqrt{3}}{2}\left(\delta _1-1\right) \cos \sigma
   _k-\frac{1}{4} \sqrt{3} \cos 2 \sigma _k+\frac{3 \sqrt{3}}{4} \right].
}
\end{equation}
where  $h= \ell /\sqrt{3}$ for the ET configuration and  $h= \ell /\sqrt{2}$ for the IRT. It can be shown \cite{cqg2} that, due to the small eccentricity and inclination of the orbits, the solution given by the above zero and first order terms differs from the exact Keplerian solution by less than $0.03 \%$ making the method of analytical series expansion a useful basis for an analytical model of the motions of LISA.

\n
Choosing $\delta_1=5/8$, the first order (Keplerian) flexing is minimized in both ET and IRT configurations, giving, with arm lengths of order 1 Gm, an extra angle of respectively $7'$ and $4'$.
{\Tref{tabdelta1} reports some orbit indicators (flexing, breathing angles, and Doppler shifts) relative to both IRT and ET configurations for $\delta_1=0$ and $\delta_1=5/8$. The indicators  $\Delta^+$ and $\Delta^-$ represent the  difference  between the maximum and minimum(respectively) value and the {\em nominal} value of a given parameter, over the 6 years of the mission.}

\begin{table}[!]
\caption{Change of the relevant orbit indicators (arm length, breathing, Doppler modulation) for the three S/C's, for both IRT and ET configurations, in the standard $\delta_1=0$ and modified $\delta_1=5/8$ inclination.
For each indicator, nominal value, average and deviations $\Delta^+$ and $\Delta^-$  (see \Sref{firstord}), relative to the nominal value over a mission lifetime of 6 years are shown. }
\begin{center}
\resizebox{0.99\columnwidth}{!}{
 \begin{tabular}{l  rrrr rrrr}
\br
  & \multicolumn{4}{c}{ IRT }& \multicolumn{4}{c}{ET}  \\ 
 $\delta_1=0$   &  nominal & mean     & $\Delta^+$ &  $\Delta^-$   &  nominal & mean     & $\Delta^+$ &  $\Delta^-$ \\
\cmidrule(l){2-5} \cmidrule(l){6-9}

 $L_{12}$  [km]    &  10$^6$                    &� 1001333 & +5210 & -969  & 10$^6$  & 1001088& +3852&-757 \\
 $L_{23}$  [km]    &  10$^6$                    &� 1001333 & +5210 & -969  & 10$^6$  & 1001088& +3859&-752 \\
 $L_{31}$  [km]    & $\sqrt{2}$ 10$^6$  &    1416098  & +4988 & -1248 &10$^6$  & 1001088& +3852&-757\\ 
\vspace{-0.2cm}\\

$\theta_{1}$  [deg]  &45 &          &           &         & 60 & 60.00& +0.27&-0.18\\
 $\theta_{2}$  [deg] &90 & 90.00& +0.36&-0.26 & 60 & 60.00& +0.27&-0.18\\
 $\theta_{3}$  [deg] &45 &          &           &         & 60 & 60.00& +0.27&-0.18\\
\vspace{-0.2cm}\\

 $\Delta \mathbf{v}_{12}$ [m/s] & - &-0.11  & +5.00 &-5.16 &- & 0.00 &+0.87&-0.87\\
 $\Delta \mathbf{v}_{23}$ [m/s] & - &+0.16 & +5.00& -4.55& - & 0.00 &+0.87&-0.87\\
 $\Delta \mathbf{v}_{31}$ [m/s] &   &           &           &         & - & 0.00 &+0.87&-0.87\\
\vspace{0.2cm}\\

\mr
$\delta_1$=5/8    &  nominal & mean     & $\Delta^+$ &  $\Delta^-$   &  nominal & mean     & $\Delta^+$ &  $\Delta^-$\\
\cmidrule(l){2-5} \cmidrule(l){6-9}
 $L_{12}$  [km]    &  10$^6$                    &� 999115   & +786   & -2554  & 10$^6$  & 999277& +241&-1686\\
 $L_{23}$  [km]    &  10$^6$                    &� 999115   & +786   & -2554  & 10$^6$  & 999277& +241&-1686 \\
 $L_{31}$  [km]    & $\sqrt{2}$ 10$^6$  &   1412962   & -1251  & -1253  & 10$^6$  & 999277 &+241&-1686 \\
\vspace{-0.2cm}\\

$\theta_{1}$  [deg]  &45 &          &           &         & 60 & 60.00& +0.09&-0.09\\
 $\theta_{2}$  [deg] &90 & 90.00& +0.12 &-0.12 & 60 & 60.00& +0.09&-0.09\\
 $\theta_{3}$  [deg] &45 &          &           &         & 60 & 60.00& +0.09&-0.09\\
\vspace{-0.2cm}\\

 $\Delta \mathbf{v}_{12}$ [m/s] & -& 0.00 & +0.27& -0.27 &- & 0.00& +0.16 &-0.16\\
 $\Delta \mathbf{v}_{23}$ [m/s] & -& 0.00 & +0.27& -0.27& -& 0.00& +0.16 &-0.16\\
 $\Delta \mathbf{v}_{31}$ [m/s] &  &           &           &   &- & 0.00& +0.16 &-0.16\vspace{0.2cm}\\
 \br
\end{tabular}
}
\end{center}\label{tabdelta1}
\end{table}

\section{The Earth effect}\label{EE}
We now include the perturbation due to the Earth. \\
In the analytic approach the Earth is assumed at rest in the Hill frame. As shown in \cite{auto}, in the {\it rotating frame}, the global dynamics in the coupled fields of Sun and Earth is characterized by secular terms producing, in the long run, a drift {\it away} from the Earth: this is linear in time in the radial direction and quadratic in the tangential direction. 
\n
In order to minimize this  quadratic $y$ drift, an intuitive strategy is to choose initial conditions such that LISA is a little further out at start, approaches the Earth, reaches its minimum distance at mid mission and departs after that \cite{cqg3}.
However, other strategies can be devised that  provide better performance of the constellation. An appreciable reduction of the flexing due to the Earth tidal field is in any case possible, over a limited time span, by suitable tuning of all orbital parameters \cite{hu,xia,li}. In our analytical approach, in order to keep things simple, we still use three identical orbits (apart for relative phase shifts, see \eref{zerovec}) for the 3 S/Cs of the constellation and, in addition to the above specs, we try and vary the tilt-angle $\delta_1$ and a subset of the initial {conditions}. 
We first consider a simplified model where the Earth describes a circular orbit of radius $R_0=1$AU around the Sun on the $xy$ plane \cite{cerdo3,auto}. Introducing the minimum  trailing angle ${\rm TA_0}$, taking place at time $t_0$,  the Earth coordinates {$(x_{\oplus},y_{\oplus},z_{\oplus})$ in the HCW frame} are

\begin{equation*}
x_{\oplus}=-R_0 (1-\cos ({\rm TA_0})), \;\;
y_{\oplus}=R_0 \sin ({\rm TA_0}), \;\; z_{\oplus}=0.
\end{equation*}

We consider its effect as a {\it constant} + {\it linear} force to be added to the equations of motion. We define the distance of the Earth from the origin,
$d_{\oplus}=\sqrt{x_{\oplus}^{2}+y_{\oplus}^{2}},$
and  introduce the perturbation parameter

\be
\varepsilon_{\oplus} = \frac{M_{\oplus}}{M_{\odot}} \left(\frac{R_0}{d_{\oplus}}\right)^{3},\quad
\frac{M_{\oplus}}{M_{\odot}} = \frac1{328900}
\label{pertpar}
\ee

so that the effect of the Earth is given by the perturbation

\begin{equation*}
\eqalign{
f_{\oplus x}  = \varepsilon_{\oplus} \omega^{2} (x_{\oplus} + C_{11} x + C_{12} y) \cr
f_{\oplus y}  = \varepsilon_{\oplus} \omega^{2} (y_{\oplus} + C_{12} x + C_{22} y)\cr
f_{\oplus z}  = \varepsilon_{\oplus} \omega^{2} (z_{\oplus} - z),}
\end{equation*}
where

\be
\label{Cij}
C_{11} = \frac{2 x_{\oplus}^2 - y_{\oplus}^2}{d_{\oplus}^2}, \;\;
C_{12} = \frac{3 x_{\oplus}y_{\oplus}}{d_{\oplus}^2}, \;\;
C_{22} = \frac{2 y_{\oplus}^2 - x_{\oplus}^2}{d_{\oplus}^2}.\ee

\n
The equations of motion can be solved with the perturbation method and the solutions, to be added to the zero and first order solutions,  are in the form

\begin{equation}
\hspace{-2cm}
\label{rE}
\eqalign{
x^{(E)}_{k}(t)=(A_{k,x}+B_{k,x} t) \sin \omega t + (C_{k,x}+D_{k,x} t) \cos \omega t + E_{k,x} + F_{x} t,\cr
y^{(E)}_{k}(t)=(A_{k,y}+B_{k,y} t) \sin \omega t + (C_{k,y}+D_{k,y} t) \cos \omega t + E_{k,y} + F_{k,y} t + G_{y} t^2,\cr
z^{(E)}_{k}(t)=(A_{k,z}+B_{k,z} t) \sin \omega t + (C_{k,z}+D_{k,z} t) \cos \omega t.
}
\end{equation}
The integration constants $A_i \cdots G_y$ are defined by the choice of initial conditions (see \ref{AppA}). The secular terms appearing in the solution generate the parabolic drift around the RV {(see \ref{SIN}).

\n
By collecting terms, the orbit of the S/C$_k$ is 
\begin{equation}\label{E012} 
\mathbf{r}_k(t)=\mathbf{r}^{(0)}_{k}(t)+\mathbf{r}^{(1)}_{k}(t)+\mathbf{r}^{(E)}_{k}(t)
\end{equation} 
where the zero and first order terms are respectively given by \eref{rzero} and \eref{runo} and $\mathbf{r}^{(E)}_{k}(t)$ is given by \eref{rE}. The terms growing as $t$ and $t^2$  in (respectively) $x^{(E)}(t)$ and  $y^{(E)}(t)$ vanish when one calculates the relative motion between S/Cs, being $F_x$ and $G_y$ equal for all S/Cs, therefore the increase in flexing with time  is only due to secular terms as $t \sin t$ and $t \cos t$. 
\n
It is now useful to define some quantities in the heliocentric frame:
\begin{itemize}
\item unit vectors of the rotating frame axes:

\begin{equation*}
\hspace{-2.5cm}
\eqalign{
\mathbf{u}_x=\{\cos (\omega t-{\rm TA}_0),\sin(\omega t-{\rm TA}_0),0\},\cr
\mathbf{u}_y=\{-\sin (\omega t-{\rm TA}_0),\cos(\omega t-{\rm TA}_0),0\},\cr
\mathbf{u}_z=\{0,0,1\}}
\end{equation*}
\item Position of the Earth: $\mathbf{R}_{\oplus}(t)=R_0\{\cos \omega t,\sin \omega t,0\}$. 
\item Orbit of S/C$_k$: 
\be\label{inertial}
\hspace{-2.5cm}
 \mathbf{R}_k(t) = (R_0+x_k(t))\mathbf{u}_x+y_k(t)\mathbf{u}_y+z_k(t)\mathbf{u}_z
\ee
\item LISA barycenter position: $\quad \mathbf{R}_g(t)= \frac13 \sum_k \mathbf{R}_k(t)$.
\item LISA arm vectors: $\quad \mathbf{R}_{ij}(t)= \mathbf{r}_{ij}(t)=\mathbf{r}_j(t)-\mathbf{r}_i(t)$.
\item LISA arm lengths: $\quad  L_{ij}(t)= \vert \mathbf{r}_{ij}(t) \vert = \vert \mathbf{R}_j(t)-\mathbf{R}_i(t) \vert $.
\item Doppler shifts: $v_{ij}(t)= \frac{d}{dt} L_{ij}(t)$
\item Trailing angle:
\[
{\rm TA} (t)=\frac{180}{ \pi} \arccos \left(\frac{\mathbf{R}_{\oplus}(t) \cdot \mathbf{R}_g(t)}{R_0\, R_g(t)} \right).
\]
\item {Breathing angles}:
\[
 \theta_{j}(t) = \frac{180}{ \pi} \arccos \left( \frac{\mathbf{r}_{ij}(t)\cdot \mathbf{r}_{jk}(t)}{L_{ij}(t) L_{jk}(t)}\right)
\]
\end{itemize}

\section{Minimization of the flexing}\label{MIN}
We shall address the choice of an orbit that minimizes the flexing of the arms in three steps:
first, by optimizing with respect to the tilt angle only. Then, by perturbing the initial conditions of the three S/Cs in a still analytic approach. Finally, with a fully numerical integration of the S/C orbits, taking into account all major perturbing effects. 
The mission begins at the time $t_{ini}$ and is assumed to last $ \Delta t =6$ years.   We refer to "mid-mission"  or $t_{mid} = t_{ini} + 3$ yrs the time half way into the mission. 

\n
In the panels of \Fref{Doppler} we consider both the ET (red) and the IRT (blue) configurations: for these plots the optimization is only done by evaluating the
optimal tilt-angle over the mission lifetime of 6 years. Optimization is performed by minimizing the RMS flexing of the 3 arms over the entire mission duration. The time of closest approach to the Earth is 3 years after the mission starts and this identifies ${\rm TA}_0$.

\begin{figure}[h!]
\centering
\includegraphics[width=0.49\columnwidth]{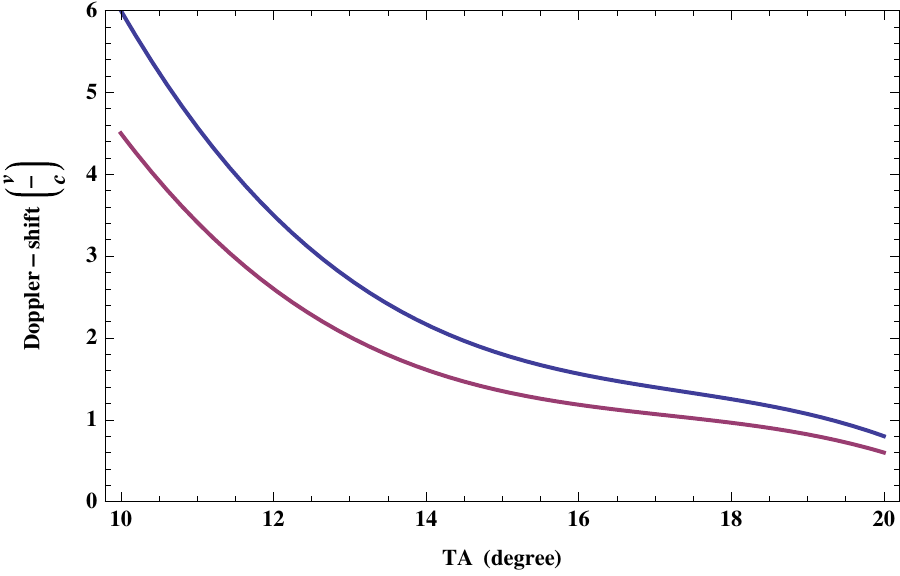}
\includegraphics[width=0.45\columnwidth]{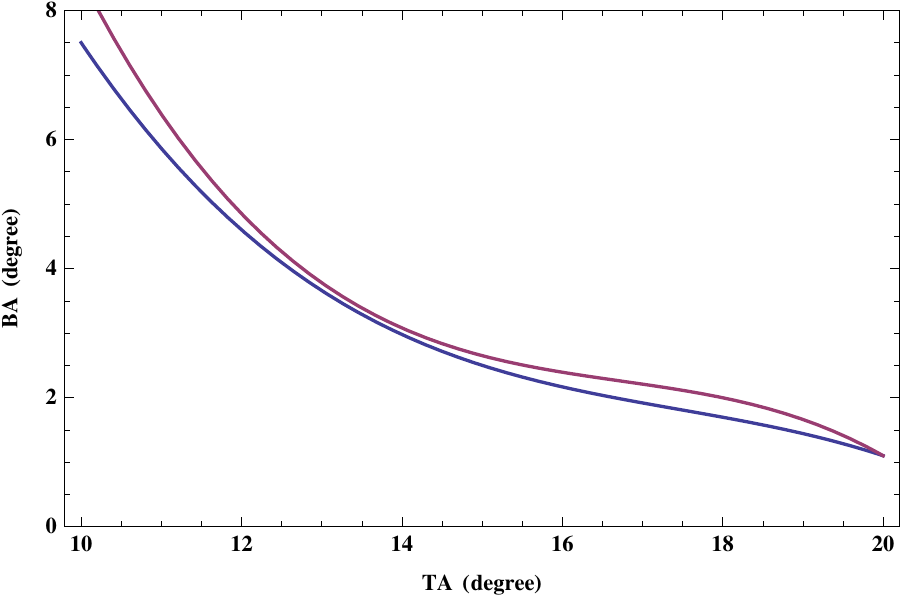}
\caption{{ Effect of the Earth (non optimized): left: maximum Doppler shift in  $m/s$. 
Right: maximum change of the angle between $L_{12}$ and $L_{13}$ (short arms in the IRT).  Red curves: ET, blue curves: IRT.
}}
\label{Doppler}
\end{figure}

\n
The requirement of $6.5 \times 10^{-8}$ on the Doppler shift shows that we can reduce the minimum TA to less than 10$^\circ$ for both cases (8$^\circ$ for the IRT). By choosing an optimal tilt angle we can strongly reduce the rms flexing of the arms; this angle always turns out to slightly differ from the canonical 60$^\circ$.
On the other hand, the breathing angle requirement of $\pm 1.5^{\circ}$ sets a limit at TA $\simeq 14^{\circ} $. Breathing appears therefore the major obstacle to an appreciable reduction of TA. 
In the next subsection, we attempt  a more general optimization strategy.

\subsection{Cost function}
In order to extend the optimization to a wider set of parameters, we need to introduce a ``cost function" , i.e. a suitable function of the relevant quantities (the flexings) that has to be minimized. 
To this purpose, we define the following cost function, suitable for both configurations (and therefore different from those proposed in \cite{hu}):
\be\label{CF1}
\hspace{-2cm}
\sigma^2=\langle(L_{12}-\langle L_{12}\rangle)^2+(L_{23}-\langle L_{23}\rangle)^2+(L_{31}-\langle L_{31}\rangle)^2 \rangle
\ee
where
\[
 \langle\dots\rangle = \frac{1}{\Delta t}\int_{t_{ini}}^{t_{fin}} \dots  dt.
\]
indicates average over the mission time.  Although in the IRT case the third arm is not monitored and its flexing could appear as a useless burden to the cost function,  we maintained the same $\sigma^2$, as defined in \eref{CF1},  for both configurations: minimizing the flexing of all arms is a way to render the triangle "more rigid" and is an effective strategy, as we shall show,  to minimize the angle breathing as well.

\subsection{Perturbation of initial conditions - semi-analytic approach}\label{semianaly}
As we extend our optimization strategy, remaining as close as possible to an analytic approach, we must restrict the space of free parameters. Our choice is a subset of the initial conditions of the unperturbed orbits. 
The Earth {produces linear and secular terms in the orbits, as shown by \eref{rE}. On the other hand, \eref{soleqhcw} show that, in general, a linear drift exists in the $y$ component (that we canceled by setting $2 \omega x_0 + \dot y_0 = 0$). We can therefore choose a suitable offset in the initial conditions in such a way that these two linear drifts compensate each other.
\n
 Therefore, if we set }  
\be\label{shift}
x_{0,k}=-\frac{\dot y_{0,k}}{2\,\omega} + \epsilon_k, \quad k=1,2,3
\ee
{the first two components of position vector \eref{zerovec} become}

\be
\eqalign{
x^{(0)}_{k} = \frac{\ell}{2 \sin (\pi / n)} \frac{1}{2} \cos \sigma_k + \epsilon_k(4 - 3 \cos \omega t),\cr
y^{(0)}_{k} = \frac{\ell}{2 \sin (\pi / n)} \sin \sigma_k - 6 \epsilon_k (\omega t - \sin \omega  t)}\ee

\n
while the third one is unchanged.

\n
 In this way trajectories and arm-lengths are affected by a perturbation which grows linearly in time. In the expansion \eref{E012} we modify only the zero-order terms: in principle, the variation should be propagated through the higher-order terms, but the contribution is of order $ \epsilon \ell / R$ in the first-order terms and even smaller in the term describing the Earth effect:  we can therefore safely neglect them.

\n
The required amount of this variation can be determined by minimizing the cost function $\sigma^2(\epsilon_1, \epsilon_2, \epsilon_3,\delta_1)$ defined in \eref{CF1}. 
However, the analytic expression for $\sigma^2$ is sufficiently cumbersome {to impose a numerical minimization:} this, on the other hand,  allows us to use the exact equations of motion:

\be\label{HCWexact1}
\eqalign{
 \ddot x_k -2\omega \dot y_k -\omega^2 (x_k+ R_0) =f_{x,k}\cr
 \ddot y_k +2\omega \dot x_k -\omega^2 y=f_{y,k}\cr
 \ddot z_k= f_{z,k}
}
\ee
where $\mathbf{f}_k$ is the Sun+Earth force per unit mass acting on the $k$-th  S/C, expressed in the HCW coordinate system.
\n
Setting the RV at $t_0=t_{mid}$, for the IRT and ET configurations respectively, the minima correspond to
\n

\be\label{OPTNAIF}
\hspace{-2cm}
\eqalign{
\rm{IRT}: \ \delta_1 = 0.808, \quad \epsilon_1 = 867\,km,\quad \epsilon_2=519\,km, \quad \epsilon_3= 66\,km,\cr
\rm{ET}: \hskip 0.3 cm \delta_1= 0.894, \quad \epsilon_1 =523\,km, \quad  \epsilon_2=64 \,km, \quad \epsilon_3= 7\,km.
}
\ee
The results of the optimization are reported  in \Tref{tabOPTNAIF}. The trailing angles in both cases at $t_{ini}$ and $t_{fin}$ are 12.8$^\circ$ degrees (33 Gm from the Earth). {The improvement  in the values of the performance indicators in the optimized cases is quite evident.}

\begin{table}[!]
\caption{
Variation of the same orbital indicators as in \Tref{tabdelta1} (arm-length, breathing, Doppler modulation) including the Earth effect (assumed on a circular orbit) 
corresponding to the optimal data of \eref{OPTNAIF} for the IRT and ET constellations (left and right, respectively).} 

\begin{center}
\resizebox{0.99\columnwidth}{!}{
 \begin{tabular}{l rrrr rrrr}
\br
   & \multicolumn{4}{c}{ IRT }& \multicolumn{4}{c}{ET}  \\

not opt.          &  nominal & mean     & $\Delta^+$ &  $\Delta^-$   &  nominal & mean     & $\Delta^+$ &  $\Delta^-$\\
\cmidrule(l){2-5} \cmidrule(l){6-9}

 $L_{12}$  [km]    &  10$^6$                    &� 1004681  & +47356 & -32740  & 10$^6$  & 1005887& +54846&-36195 \\
 $L_{23}$  [km]    &  10$^6$                    &� 1005025  & +55742 & -42716  & 10$^6$  & 1001830& +16574&-16501 \\
 $L_{31}$  [km]    & $\sqrt{2}$ 10$^6$  &   1425000   & +92213  & -59335  & 10$^6$  & 1006176 &+59722 &-40762 \\
\vspace{-0.2cm}\\
 
$\theta_{1}$  [deg]  &  &          &           &         & 60 & 59.73& +2.71&-3.50\\
 $\theta_{2}$  [deg] &90 & 90.30& +4.27&-2.89 & 60 & 60.14& +3.91&-3.48\\
 $\theta_{3}$  [deg] &  &          &           &         & 60 & 60.11& +4.00&-3.48\\
\vspace{-0.2cm}\\

 $\Delta \mathbf{v}_{12}$ [m/s] & -&-0.36 & +8.76&-12.15 & & -0.30 & +10.93&-13.95\\
 $\Delta \mathbf{v}_{23}$ [m/s] & -&+0.49 & +13.07& -10.17& &+0.06& +5.62 &-5.04\\
 $\Delta \mathbf{v}_{31}$ [m/s] &  &  &           &         & &+0.43 & +14.72&-12.79\\
\vspace{0.2cm}\\
\mr
optimized  &  nominal & mean     & $\Delta^+$ &  $\Delta^-$   &  nominal & mean     & $\Delta^+$ &  $\Delta^-$\\

\cmidrule(l){2-5} \cmidrule(l){6-9}
 $L_{12}$  [km]    &  10$^6$                    &� 999363  & +14322 & -16976  & 10$^6$  & 999440& +13569&-16262 \\
 $L_{23}$  [km]    &  10$^6$                    &� 999284  & +14665 & -15256  & 10$^6$  & 999228& +12924&-15261 \\
 $L_{31}$  [km]    & $\sqrt{2}$ 10$^6$  &   1413390  & +18695 & -21444  &10$^6$  & 999353&+12793 &-15472 \\
\vspace{-0.2cm}\\

 $\theta_{1}$  [deg]  &  &          &           &            & 60 & 59.99& +1.15&-1.16\\
 $\theta_{2}$  [deg] &90 & 90.01& +1.48&-1.50 & 60 & 60.00& +1.19&-1.24\\
 $\theta_{3}$  [deg] &  &          &           &            & 60 & 60.01& +1.27&-1.26\\
\vspace{-0.2cm}\\

 $\Delta \mathbf{v}_{12}$ [m/s] & -&-0.11 & +5.00 & -5.16 &-& -0.08 & +4.88&-5.14\\
 $\Delta \mathbf{v}_{23}$ [m/s] & -&+0.16 & +5.00& -4.55& -&+0.01& +5.02 &-5.05\\
 $\Delta \mathbf{v}_{31}$ [m/s] &  &  &           &                  & -&+0.14 & +4.97&-4.69\\
 \br
\end{tabular}
}
\end{center}\label{tabOPTNAIF}
\end{table}

\subsection{Numerical optimization }\label{numeric}
In this section we describe the fully numeric evaluation and minimization of the cost function \eref{CF1} by solving the exact equations of motion and taking into account 
perturbing effect  of the Sun, Venus, Earth, Moon, Mars and Jupiter. Their real trajectories
$\mathcal{R}_\odot(t),  \mathcal{R}_{\oplus} (t), \mathcal{R}_{\rightmoon}(t), \mathcal{R}_{\venus}(t), \mathcal{R}_{\mars}(t), \mathcal{R}_{\jupiter}(t)$
in the Solar System Barycenter (SSB), are provided by the JPL HORIZON ephemerides \cite{JPL}, with the following characteristics:

\begin{itemize}
\item    Reference epoch: J2000.0
\item    $XY$-plane: plane of the Earth's orbit at the reference epoch.
\item    $X$-axis: out along ascending node of instantaneous plane of the Earth's orbit and the Earth's mean equator at the reference epoch.
\item    $Z$-axis: perpendicular to the XY-plane in the directional (+ or -) sense of Earth's north pole at the reference epoch.
\item step: 1 day.
\end{itemize}
\n
In the simplified model used till here, the Sun is assumed at rest at the center of an inertial  frame. However, the true inertial frame is represented by the Solar System Barycenter (SSB), where the Sun moves in a non negligible and complex {(non simply periodic)} way: in this frame the motion of the Earth  is substantially different from an ellipse, and therefore the initial condition that we adopted  for the S/Cs using \eref{soleqhcw} are no longer suitable.  Moreover, the motion of the Sun is also a relevant source of perturbation.  
Therefore, to account for these additional effects while maintaining the convenient, Sun-centered HCW  description, we must complete the equations of motion with an apparent force deriving from the acceleration of the Sun relative to the SBB.
\n
The equations of motion are as \eref{HCWexact1}, with the forcing term modified as follows

\[
\hspace{-2cm}
f_{x,k}=(\mathbf{f}_k-\ddot \mathcal{R}_\odot)\cdot \mathbf{u}_x, \quad f_{y,k}=(\mathbf{f}_k-\ddot \mathcal{R}_\odot)\cdot \mathbf{u}_y \quad f_{z,k}=(\mathbf{f}_k-\ddot \mathcal{R}_\odot)\cdot \mathbf{u}_z.
\]
\n
where $\mathbf{f}_k$ is the total Newtonian force per unit mass on the $k$-th S/C.
\begin{equation}
\hspace{-2cm}
\mathbf{f}_k=-\sum_\alpha \frac{G M_\alpha}{\Vert \mathcal{R}_\odot-\mathcal{R}_\alpha+\mathbf{R}_k\Vert^3}(\mathcal{R}_\odot-\mathcal{R}_\alpha+\mathbf{R}_k),    
\quad \alpha= \odot, \venus, \oplus, \rightmoon, \mars, \jupiter .
\end{equation}
\n
and  $\mathbf{R}_k$ is the position of  $k$-th S/C in the heliocentric frame (given by \eref{inertial}).

\n
{The amplitude of flexing and breathing scales inversely with the LISA-Earth distance. This can be intuitively explained as follows: a small flexing is obtained if the constellation rapidly moves away from the Earth, its main source of perturbation to a rigid configuration. However, the overall distance in the mission lifetime must be bound within reasonable values dictated by communication requirements.
\n
An analytical study of the evolution of the Earth-LISA distance is shown in \ref{SIN} were it is verified that the LISA-Earth distance increases as $t^2$, after (and before) the RV. Moreover, there is an additional sinusoidal modulation at 1 year period due to  the eccentricity of the Earth's orbit.
We prove that the minima of the sinusoid occur at well defined epochs that depend on the allowed minimum TA but not on the epoch $t_0$ of the RV.
Therefore, in order to minimize the Earth-LISA distance, the optimal choice for $t_0$ is just one of these minima \eref{t0}. 
This shows,  as mentioned in \Sref{EE},  that other choices of $t_0$,  different from $t_{mid}$, can minimize flexing and breathing. 
\n
 In the following we shall discuss two cases:  $t_0 = t_{ini}$ and $t_0 = t_{mid}$.
The value of TA$_0$ is chosen as the minimum one that allows a breathing angle smaller that $1.5^\circ$, as required.
For the two  configurations and the two kinds of RV considered the minima of the cost function is found at the following values of parameters:
\n
IRT - RV at the beginning, ($t_0=  t_{ini}$):
\[
\hspace{-2cm}
\qquad  \delta_1=0.061, \quad \epsilon_1=430\,km, \quad  \epsilon_2=-113\,km, \quad \epsilon_3=-9\,km.
 \]
\n
IRT - RV at mid mission ($t_0= t_{mid}$):
\[
\hspace{-2cm}
\qquad  \delta_1=-0.290, \quad \epsilon_1=28\,km, \quad  \epsilon_2=-55\,km, \quad \epsilon_3=-170\,km.
\]
 \n
ET - RV at the beginning ($t_0=  t_{ini}$):
\[
\hspace{-2cm}
\qquad  \delta_1=0.473, \quad \epsilon_1=70\,km, \quad  \epsilon_2=-483\,km, \quad \epsilon_3=42\,km.
 \]
\n
ET - RV at mid mission ($t_0= t_{mid}$):
\[
\hspace{-2cm}
\qquad \delta_1=-0.047 \quad \epsilon_1=193\,km, \quad  \epsilon_2=52\,km, \quad \epsilon_3=18\,km.
\]

\n
\Tref{tabOPTJPL} provides more results and details  for  the four  cases (2 configurations $\times$ 2 RV times) considered here.
\n
 Some results are also plotted in  \Fref{OPTJPLIRT} and \ref{OPTJPLET} for IRT and ET configurations, respectively. The ranges of LISA-Earth distances and trailing angles, as well as the initial conditions for the S/Cs are reported in \Tref{NumIC}. 
We observe that the minimum TA is larger when the RV is at mid mission, but the $\Delta$TA is smaller. In general, breathing angles within the specs of $1.5^\circ$, can be obtained at smaller distance from Earth for the ET than the IRT configuration.

\begin{figure}[h!]
\centering
\includegraphics[width=0.49\columnwidth]{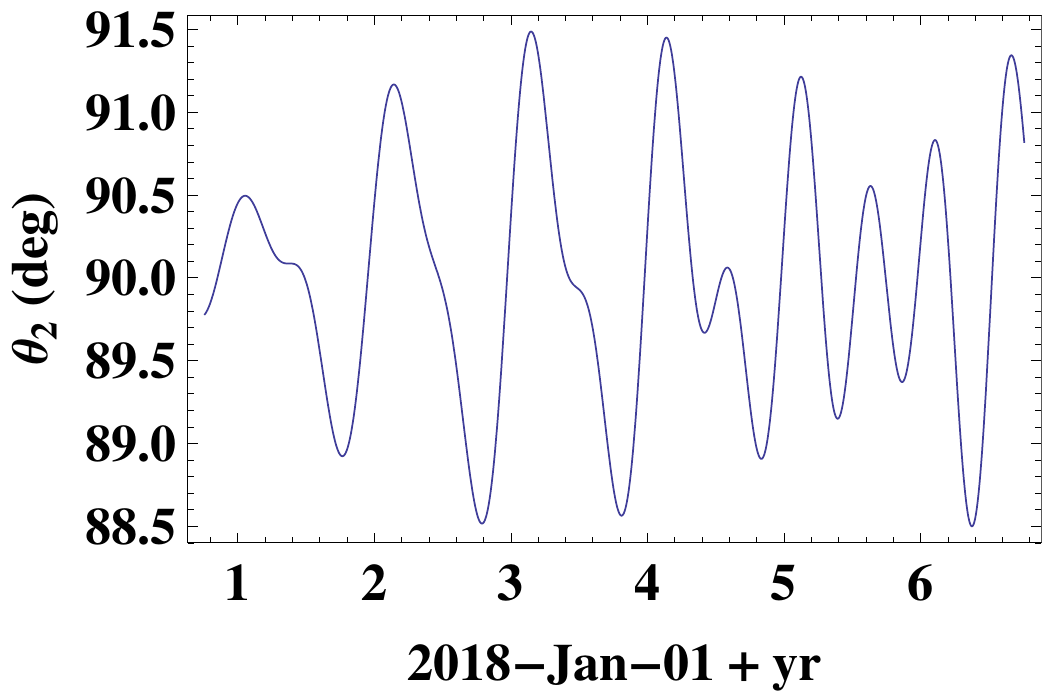}
\includegraphics[width=0.49\columnwidth]{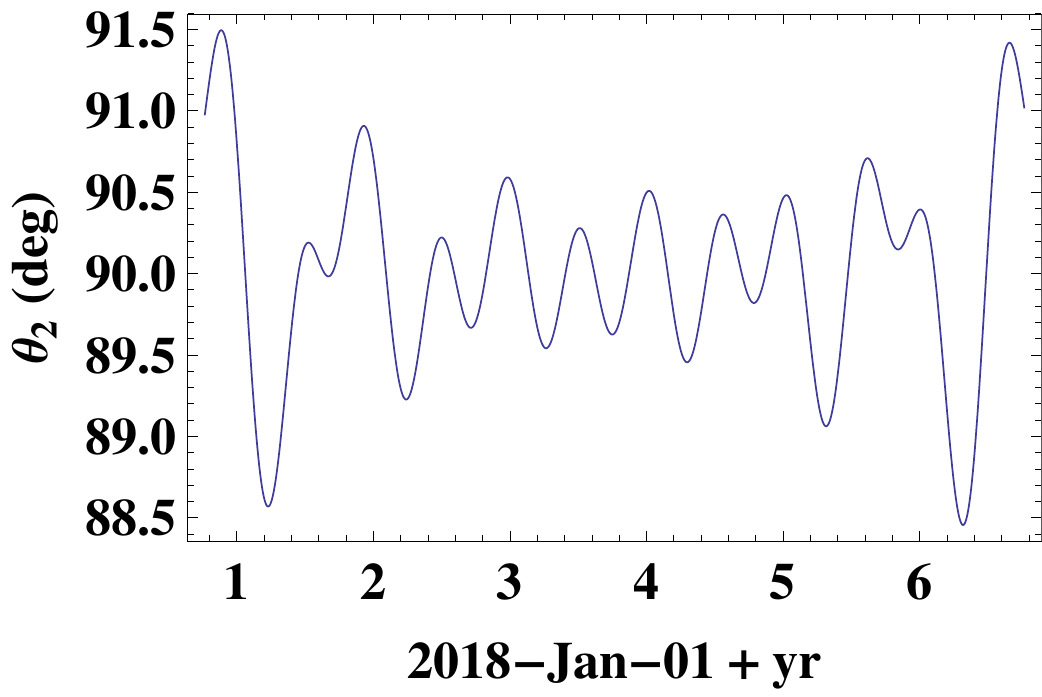}
\includegraphics[width=0.49\columnwidth]{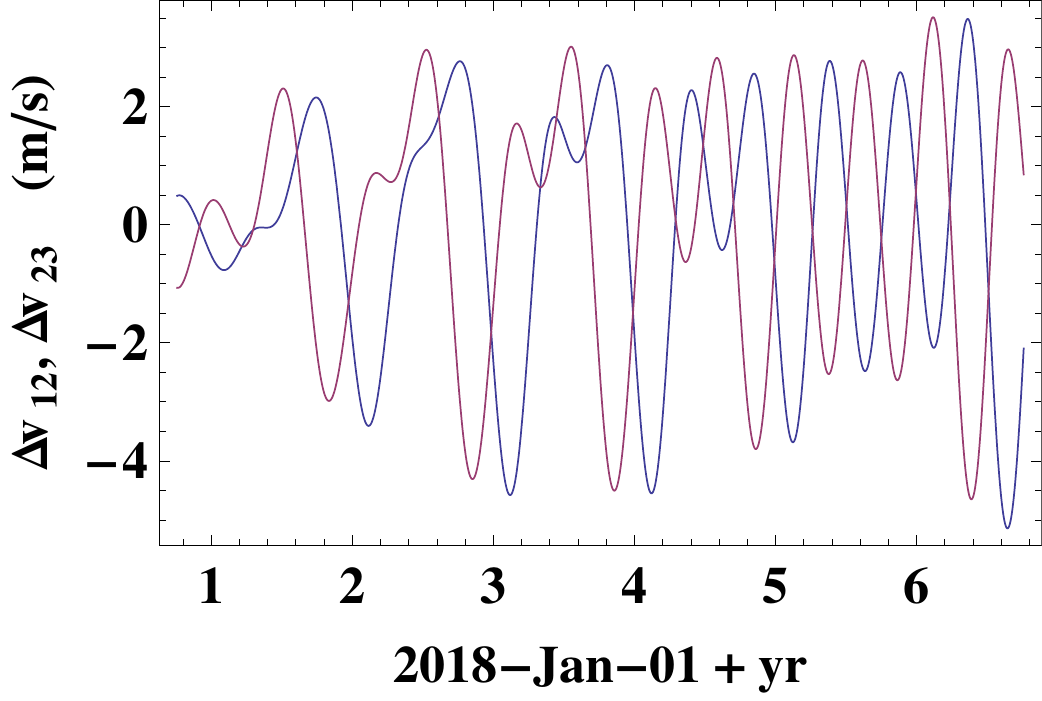}
\includegraphics[width=0.49\columnwidth]{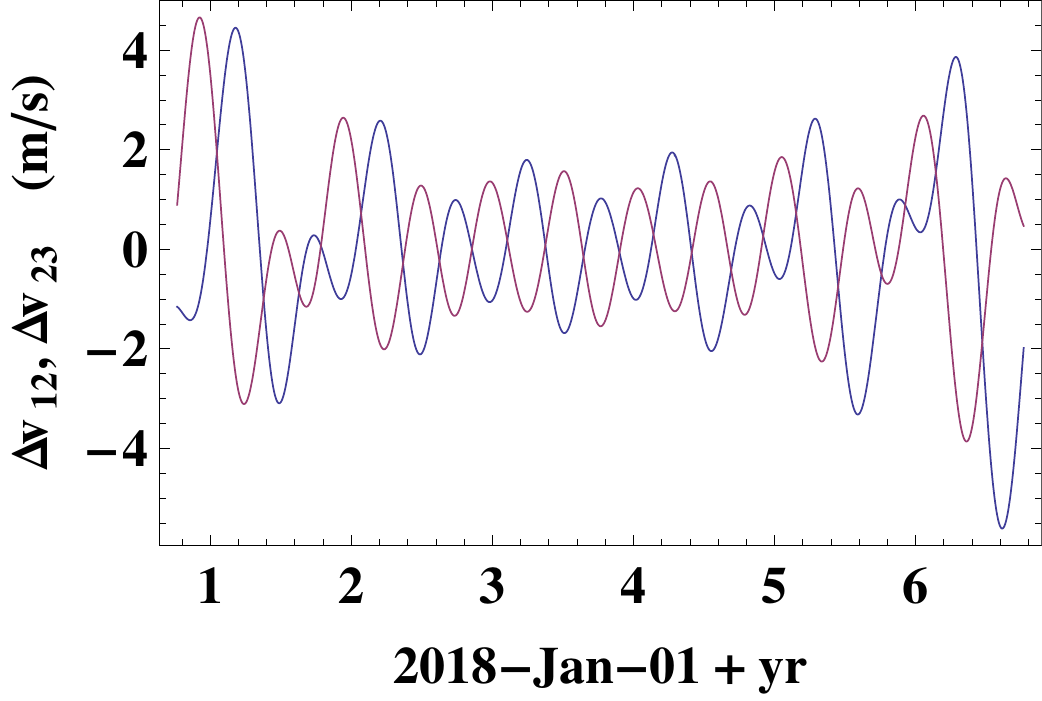}
\includegraphics[width=0.49\columnwidth]{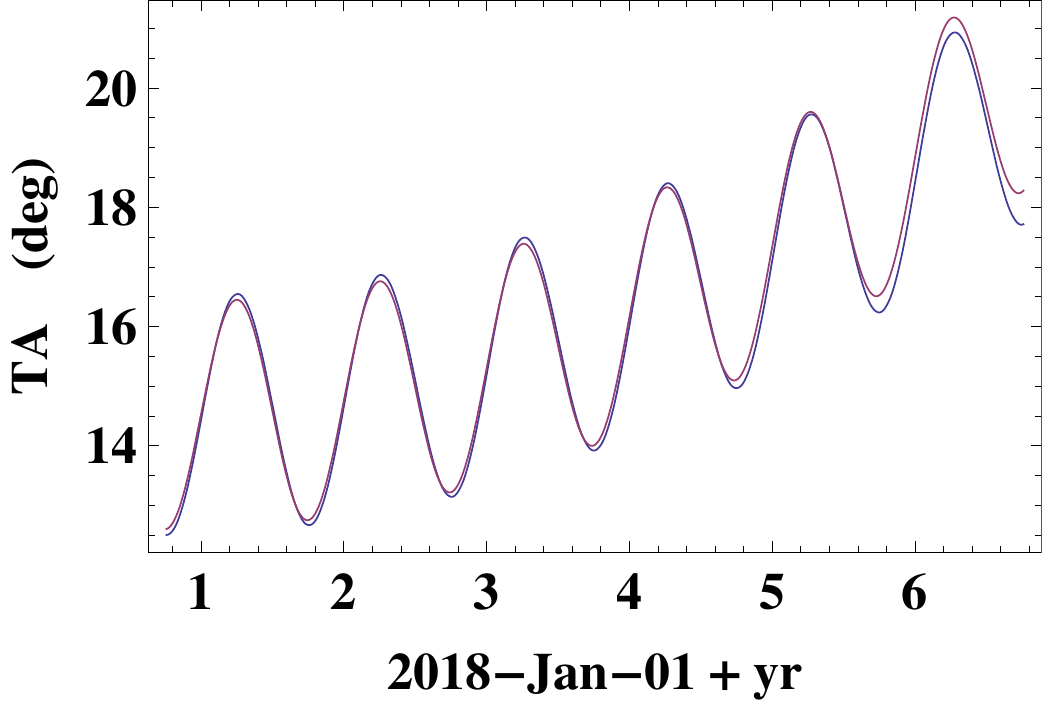}
\includegraphics[width=0.49\columnwidth]{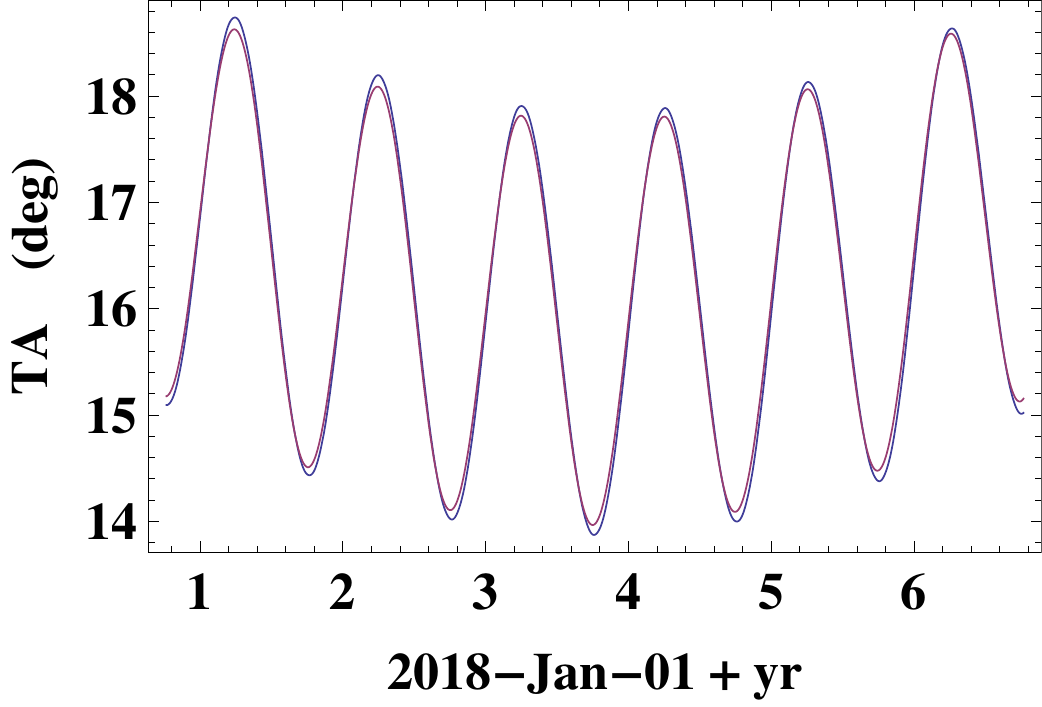}
\caption{Fully numerical optimization (\Sref{numeric}) with respect to both initial conditions and tilt angle for the IRT configuration. Left panels: RV at the beginning of the mission, right panels: RV at mid-mission. Top panels: breathing angles. Center panels: Doppler shifts. Lower panels: distance LISA-Earth, expressed as TA (red lines are obtained using \eref{last}.}
\label{OPTJPLIRT}
\end{figure}

\begin{figure}[h!]
\centering
\includegraphics[width=0.49\columnwidth]{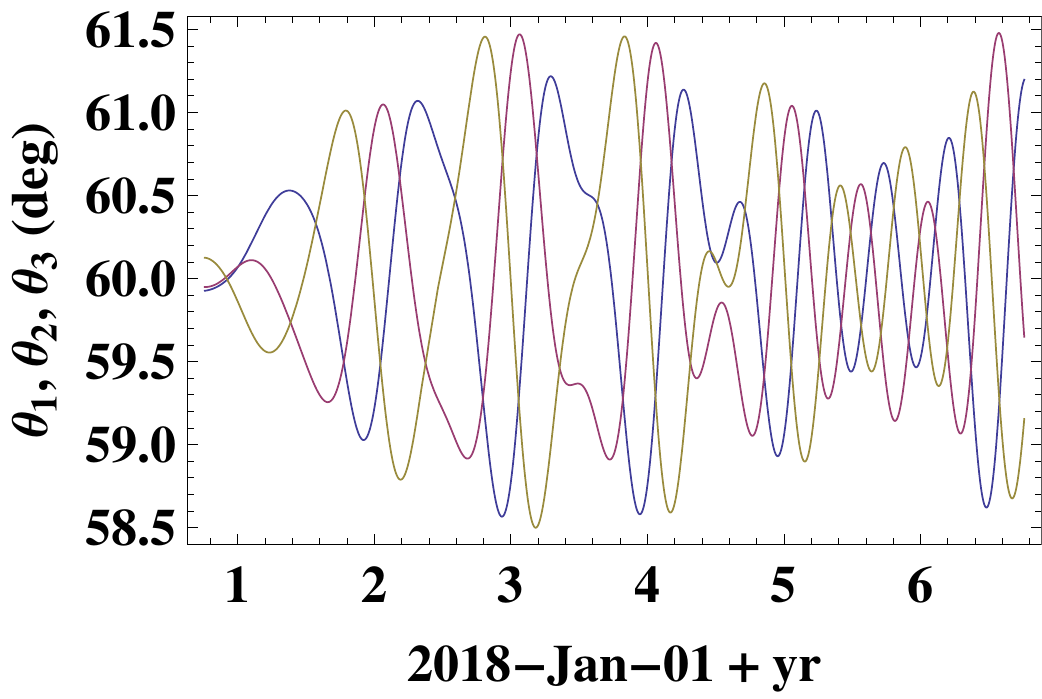}
\includegraphics[width=0.49\columnwidth]{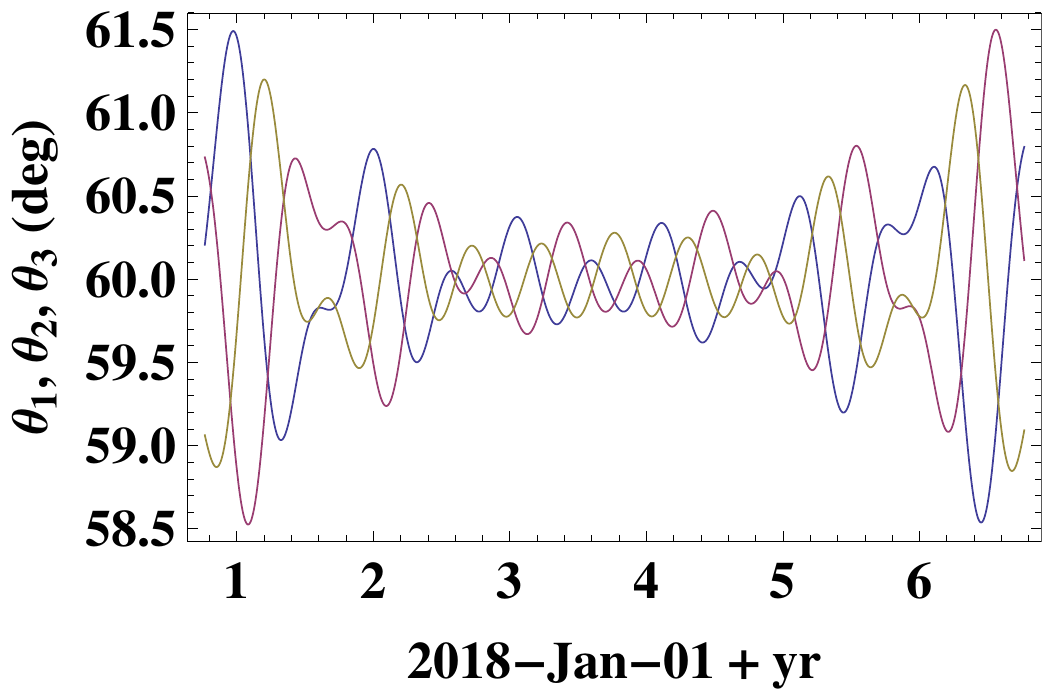}
\includegraphics[width=0.49\columnwidth]{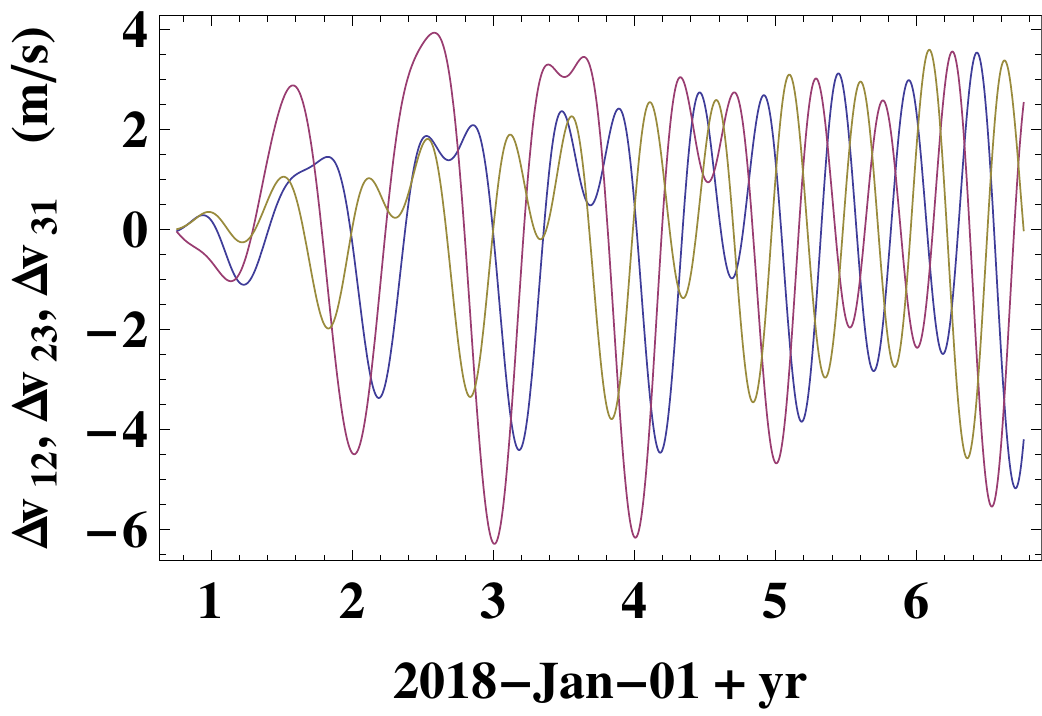}
\includegraphics[width=0.49\columnwidth]{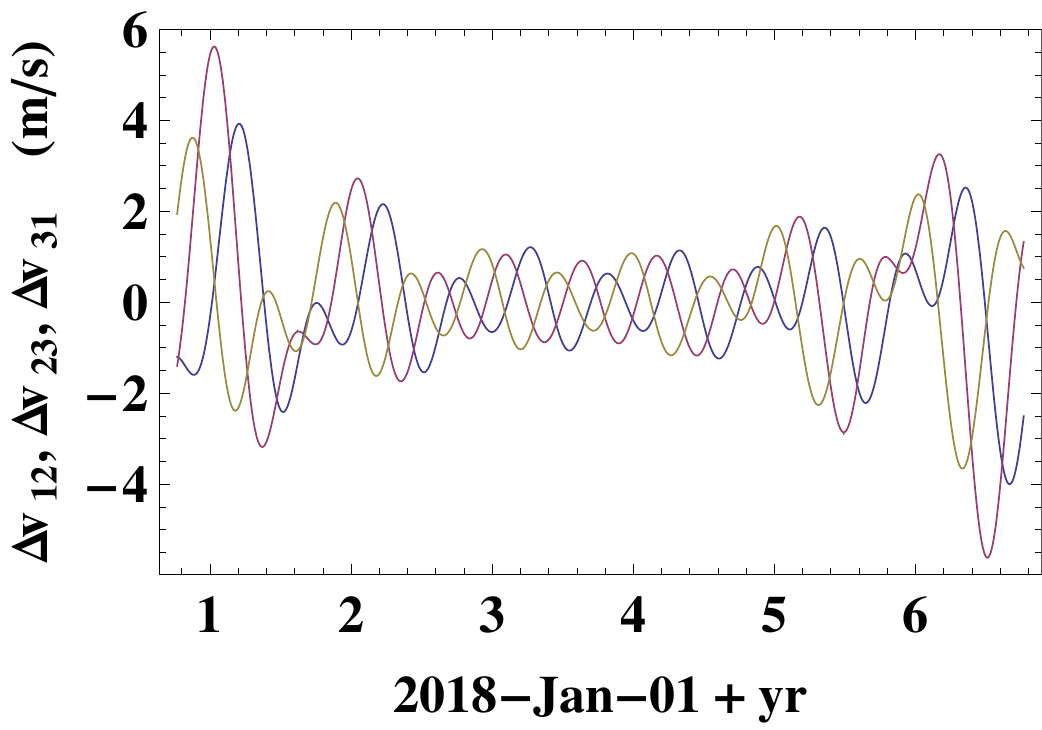}
\includegraphics[width=0.49\columnwidth]{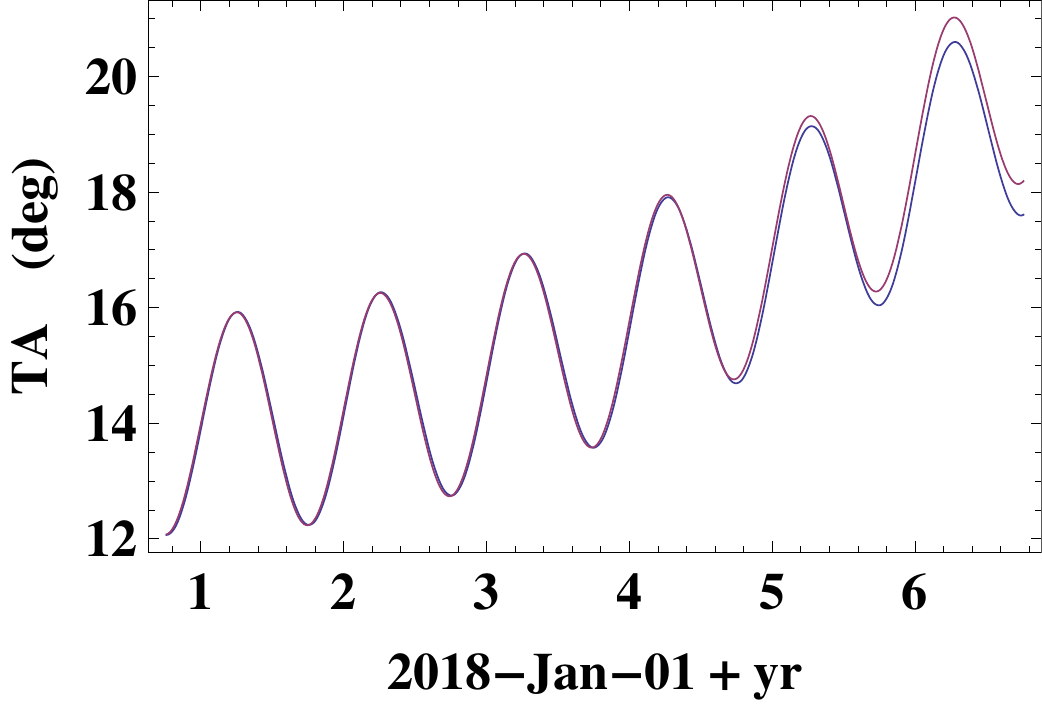}
\includegraphics[width=0.49\columnwidth]{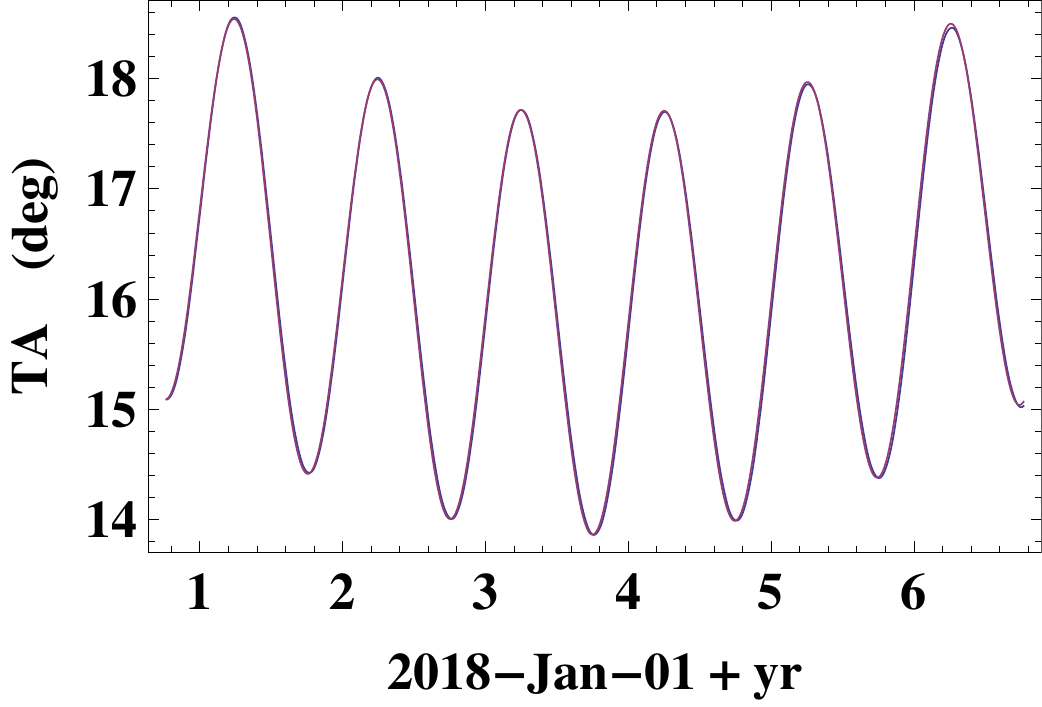}

\caption{Same as in \Fref{OPTJPLIRT}, but for the ET configuration. 
 Left panels: RV at the beginning of the mission, right panels: RV at mid-mission. Top panels: breathing angles. Center panels: Doppler shifts. Lower panels: distance LISA-Earth, expressed as TA (red lines are obtained using \eref{last}
 }
\label{OPTJPLET}
\end{figure}

\n
\Fref{finale} shows, in a synoptic way, the results of our optimization procedure with respect to Doppler and breathing angle, vs. the minimum trailing angle TA$_0$. By comparing these optimized results with those derived from the simplest model shown in \Fref{Doppler},  we see that, even considering many more perturbing agents, the optimization manages to reduce  both performance indicators by about a factor of 2  at small TA. Again we see that the requirement on the breathing remains the most stringent constraint. However, while complying with keeping the breathing within  $\pm 1.5^{\circ}$, we can address the reduction of flexing following again two opposite strategies: we can set the RV at the beginning of the mission, achieving the lowest values  of  TA$_0$  (we have $12.5^{\circ}$ for the IRT and $12.1^{\circ}$  for the ET), and accept a maximum $\Delta$TA of about 8.5 degrees in both cases. 
Else, if RV takes place at mid-mission, we must accept larger  values of TA$_0$ ($13.9^{\circ}$ for the IRT and $13.8^{\circ}$  for the ET), but TA will change much less during the mission: $\Delta$TA is less than 4.8 degrees in both cases.

\begin{figure}[h!]
\centering
\includegraphics[width=0.49\columnwidth]{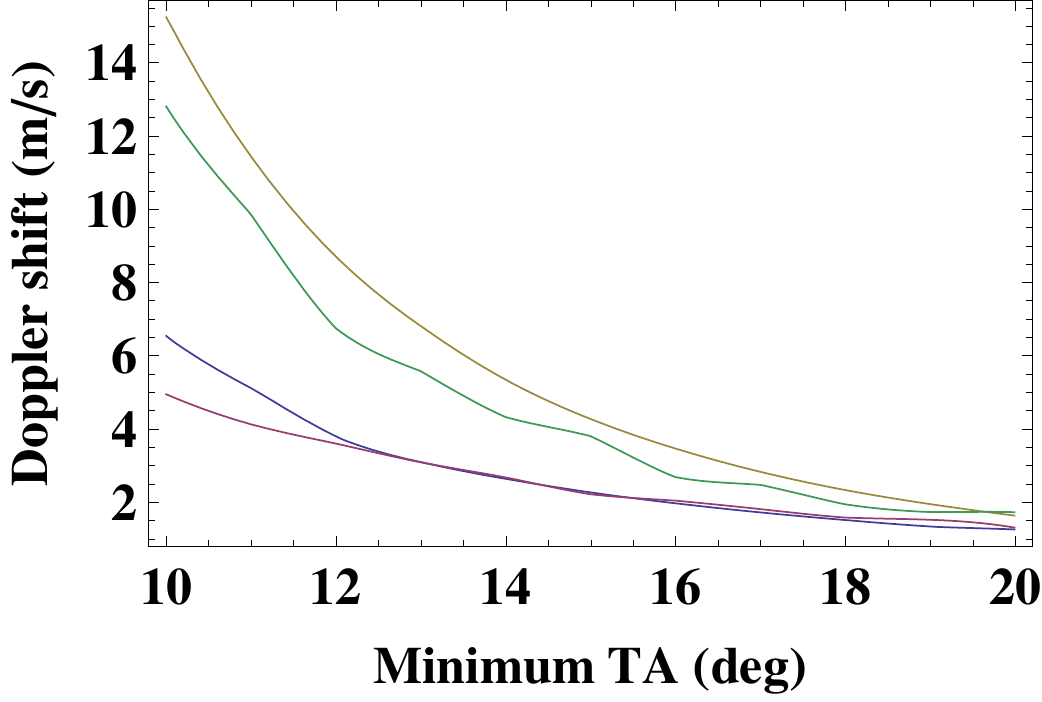}
\includegraphics[width=0.48\columnwidth]{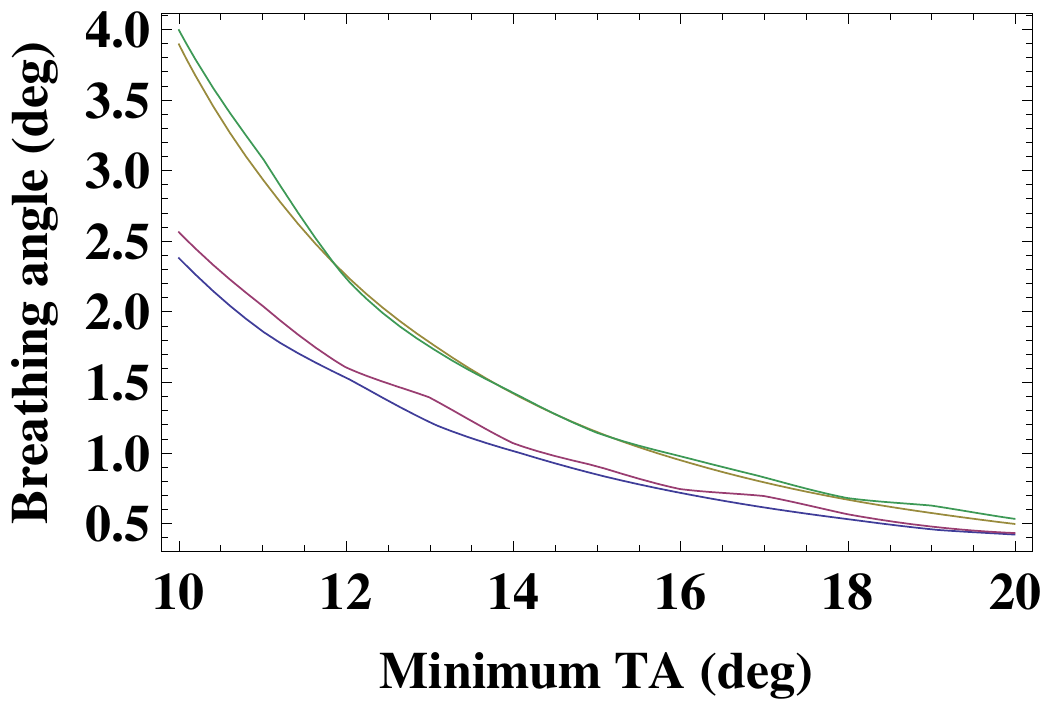}
\caption{Results of the optimization.  Left: maximum Doppler shift in  $m/s$. 
Right: maximum change of the angle between $L_{12}$ and $L_{13}$ (short arms in the IRT).  Red curves: ET (RV at the beginning of the mission), blue curves: IRT (RV at the beginning of the mission), yellow curves: ET (RV at mid mission), green curves: IRT (RV at mid mission).}
\label{finale}
\end{figure}

\begin{table}
\caption{Variation of the same orbital indicators as in \Tref{tabdelta1} and \ref{tabOPTNAIF} (arm length, breathing, Doppler modulation) including the effect of the main bodies of the Solar System  for the IRT and ET constellations (left and right, respectively).}
\begin{center}
\resizebox{0.99\columnwidth}{!}{
 \begin{tabular}{l rrrr  rrrr}
\br
  & \multicolumn{4}{c}{ IRT }& \multicolumn{4}{c}{ET}  \\

RV at $t_{ini}$           &  nominal & mean     & $\Delta^+$ &  $\Delta^-$   &  nominal & mean     & $\Delta^+$ &  $\Delta^-$ \\
\cmidrule(l){2-5} \cmidrule(l){6-9}

 $L_{12}$  [km]    &  10$^6$                    &� 1001798  & +19402 & -13718   & 10$^6$  & 1000466 & +15507&-15009 \\
 $L_{23}$  [km]    &  10$^6$                    &� 1002801  & +20096 & -11128   & 10$^6$  & 1001006 & +27444&-24115 \\
 $L_{31}$  [km]    & $\sqrt{2}$ 10$^6$  &    1417361   & +17162 & �-9388  & 10$^6$ & 1001529 &+14365&-11594\\
\vspace{-0.2cm}\\

$\theta_{1}$  [deg]  &   &          &           &         & 60 & 59.99& +1.45&-1.48\\
 $\theta_{2}$  [deg] &90 & 89.99& +1.48&-1.48 & 60 & 60.05& +1.21&-1.42\\
 $\theta_{3}$  [deg] &  &          &           &         & 60 & 59.94& +1.47&-1.08\\
\vspace{-0.2cm}\\

 $\Delta \mathbf{v}_{12}$ [m/s] & -&-0.07  & +3.35&-5.14 & - & -0.05 & +3.32&-5.17\\
 $\Delta \mathbf{v}_{23}$ [m/s] & -&+0.01 & +3.44& -4.62& - &-0.08 & +3.91 &-6.15\\
 $\Delta \mathbf{v}_{31}$ [m/s] &  &           &           &         & - &+0.03 & +3.57&-4.46\\
\vspace{-0.2cm}\\

distance [Gm] & \multicolumn{4}{c}{32.6 $\div$ 54.5}& \multicolumn{4}{c}{31.4 $\div$ 53.6}\\
TA [deg]&\multicolumn{4}{c}{12.5 $\div$ 21.0}& \multicolumn{4}{c}{12.1 $\div$ 20.7} \vspace{0.4cm}\\

\mr
RV at $t_{mid}$  &  nominal & mean     & $\Delta^+$ &  $\Delta^-$   &  nominal & mean     & $\Delta^+$ &  $\Delta^-$ \\

\cmidrule(l){2-5} \cmidrule(l){6-9}
 $L_{12}$  [km]    &  10$^6$                    &� 1002735  & +25125 & -15193  & 10$^6$  & 1001438& +16132&-14048 \\
 $L_{23}$  [km]    &  10$^6$                    &� 1002580  & +17867 & -13678  & 10$^6$  & 1001331& +22290&-22441 \\
 $L_{31}$  [km]    & $\sqrt{2}$ 10$^6$  &   1418240   & +15529 & �-3942 &10$^6$  & 1001359&+14976&-10561\\
 \vspace{-0.2cm}\\

$\theta_{1}$  [deg]  &  &          &           &         & 60 & 59.99& +1.18&-1.12\\
 $\theta_{2}$  [deg] &90 & 90.02& +1.49&-1.49 & 60 & 60.00& +1.45&-1.45\\
 $\theta_{3}$  [deg] &  &          &           &         & 60 & 60.00& +1.46&-1.46\\
\vspace{-0.2cm}\\

 $\Delta \mathbf{v}_{12}$ [m/s] & -&-0.04 & +4.44 & -5.44 &-& -0.02 & +3.86&-3.92\\
 $\Delta \mathbf{v}_{23}$ [m/s] & -&+0.06& +4.56& -4.82 & -&+0.01& +5.52 &-5.55\\
 $\Delta \mathbf{v}_{31}$ [m/s] &  &  &           &                  & -&+0.04 & +3.60&-3.62\\
\vspace{-0.2cm}\\

distance [Gm] & \multicolumn{4}{c}{36.1 $\div$ 48.6}& \multicolumn{4}{c}{36.1 $\div$ 48.2}\\
TA [deg]&\multicolumn{4}{c}{13.9 $\div$ 18.7}& \multicolumn{4}{c}{13.8 $\div$ 18.5}\\
\br
\end{tabular}
}
\end{center}\label{tabOPTJPL}
\end{table}

\begin{table}[!]
\caption{Initial conditions for the IRT (top) and ET (bottom) configuration in the heliocentric reference frame.
}
\begin{center}
\resizebox{0.95\columnwidth}{!}{
\begin{tabular}{l  rrrrrr}
\br
    & $X(t_0)$ & $Y(t_0)$ & $Z(t_0)$ & $\dot X(t_0)$ & $\dot Y(t_0)$ & $\dot Z(t_0)$ \\
        & [Gm]     & [km]     & [km]     & [km/h]        & [km/h]        & [km/h] \\
  \vspace{-0.2cm}     \\

IRT, RV at $t_{ini}$ &\multicolumn{6}{c}{$t_{min}$= 2018-Oct-05}\\
S/C1  &  149932288 & 2412779 & 612461 & 1721 & 106957 & 0\\
S/C2  &  149588968 & 1697856 & 2895     & 1471 &107213 & 438\\
S/C3  & 149222635 & 2401359 & 609567  & 1729 &107465 & 0 \\
\vspace{-0.2cm}     \\

IRT, RV at $t_{mid}$&\multicolumn{6}{c}{$t_{min}$= 2018-Oct-07}\\
S/C1     & 149753118 & 2563623 & 611220 & 1831 & 107004 & 10\\
S/C2     & 149404775 & 1872191 & 11110   & 1595 & 107265 & 438\\
S/C3     & 149042293 & 2568312 &608484  & 1845 & 107514 & 9\\
 \vspace{-0.2cm}     \\

ET, RV at  $t_{ini}$ &\multicolumn{6}{c}{$t_{min}$= 2018-Oct-05}\\
S/C1     & 149884804 & 553415   & 500457 & 395 & 107017 & 0\\
S/C2     & 149453230 & 51258     & 248057 & 217 & 107327 & 311 \\
S/C3     & 149450059 & 1052373 & 248057 & 576 & 107326 & 311\\
 \vspace{-0.2cm}     \\

ET, RV at $t_{mid}$ &\multicolumn{6}{c}{$t_{min}$=2018-Oct-07}\\
S/C1  & 149701044 & 1382500 & 499337 & 987  & 107064 & 9 \\
S/C2  & 149267830 & 902343   & 257543 & 824  & 107378 & 307 \\
S/C3  & 149266254 & 1881003 & 237763 &1167 & 107370 & 314 \\
\br
 \end{tabular}
}
\label{NumIC}
\end{center}
\end{table}

\section{Conclusions}\label{conclusions}

We have shown that the choice  of heliocentric orbits for LISA is a viable solution even when reducing the arm-length: this allows a substantial reduction in the TA (with deriving beneficial savings for placement in orbit and communications with Earth), of an amount that depends on the assumed mission duration. For an expected mission time of 6 years, the minimum value of the TA can be reduced to about $12^\circ$.
Should a 2-link interferometer be preferred for a new, cheaper version of the LISA mission, the Isosceles Right Triangle is a viable configuration, as stable as the Equilateral Triangle in all of the tests we have computed.
The amount of flexing that  the constellation undergoes during the mission depends strongly on the initial conditions.
 The reasons of this behaviour lie mostly in the time dependent perturbations due to the eccentricity of the Earth orbit and to the motion of the Sun with respect to the Solar System Barycenter. A more detailed analysis of these effects is underway. 

\clearpage
\appendix
\section{Initial Conditions for the motion in the Earth field.}\label{AppA}
\Eref{rE} can be recast in the following, equivalent but more explicit form, where the constants $C_{ij}$ of \eref{Cij}  are folded into the solution: 

\ba
x_{k}^{(2)} &=& x_{\oplus} + 2  A_{k} + 2 y_{\oplus} \omega t +
B_{k} \cos \omega t + C_{k} \sin \omega t + \nn\\
&& \ell \frac{(C_{11} + 12 C_{22}) \cos \sigma_{k} + 2 (2 C_{12} + (C_{11}
+ 4 C_{22}) \omega t)\sin \sigma_{k}}{8\sqrt{3}},   \nn \\
y_{k}^{(2)} &=& 4 y_{\oplus}-(3 \frac{A_{k}}{2} + 2 x_{\oplus}) \omega t -
\frac32 y_{\oplus} (\omega t)^{2} + D_{k}\nn \\
&& + 2 (C_{k} \cos \omega t  - B_{k} \sin \omega t)   \nn \\
&& -\ell  \frac{(3C_{11} + 16 C_{22}) \cos \sigma_{k} - 2 (2 C_{12} +
(C_{11} + 4 C_{22}) \omega t)\cos \sigma_{k}}{4\sqrt{3}},\nn \\
z_{k}^{(2)} &=& E_{k} \cos \sigma_{k} + F_{k} \sin  \sigma_{k} - \frac{\ell}4
 \sigma_{k} \sin  \sigma_{k}, \nn
\ea
\n
where the $\sigma_{k}$ are the time-dependent phases \eref{zerovec}.
\n
The 18 constants  $A_k,...,F_k$ are  determined by the initial conditions, that are chosen assuming ($x^{(E)}_{k},\,y^{(E)}_{k},\,z^{(E)}_{k}$)=(0,0,0) at $t=0$. 
\n
They are:

\ba
A_{k} &=& -\ell \frac{2 C_{22} \cos \sigma_{k}^0 - C_{12} \sin \sigma_{k}^0}{\sqrt{3}} ,  \nn \\
B_{k} &=& -x_{\oplus} -\frac{\sqrt{3} \ell}{24} ((C_{11}-4C_{22}) \cos \sigma_{k}^0 + 4 C_{12} \sin \sigma_{k}^0), \nn \\
C_{k} &=& - 2 y_{\oplus} +\frac{\sqrt{3} \ell}{24} ((C_{11}-4C_{22}) \sin \sigma_{k}^0 - 4 C_{12} \cos \sigma_{k}^0), \nn \\
D_{k} &=& \frac{ \ell}{2 \sqrt{3}} (C_{12} \cos \sigma_{k}^0 - 2( C_{11} + 3 C_{22}) \sin \sigma_{k}^0), \nn \\
E_{k} &=& -\frac{\ell}{4} \sin^2 \sigma_{k}^0 , \hskip 2cm
F_{k} = -\frac{\ell}{8} \left( 2 \sigma_{k}^0  - \sin 2 \sigma_{k}^0 \right), \nn 
\ea

\n
where the $\sigma_{k}^0 =\frac{2\pi (k-1)}{n} $ are the relative phase shifts of \eref{zerovec} evaluated at $t=0$.

\section{Distance Earth-LISA barycenter in epicyclic approximation}\label{SIN}
Here we calculate the analytic expression for the distance between a particle (i.e.: the LISA barycenter) and the Earth, taking into account the eccentricity of the orbit.

\n
We consider the Earth orbit in epicyclic approximation (at $t=0$ in the perihelion) in the inertial frame centered in the Sun:
\be\label{epicyclic}
\eqalign{
\mathbf{R}_\oplus(t)=R_0  \{ \cos \omega t+\frac{e}{2}\cos 2 \omega t-\frac{3}{2} e, \sin\omega t+\frac{e}{2}\sin 2 \omega t,0 \}.
}
\ee
The LISA barycenter, as a first approximation, can be considered at rest in the HCW frame at TA$_0$ degrees from the Earth. In the inertial frame its trajectory is 
\[
\mathbf{R}_g(t)=R_0\{ \cos ( \omega t-{\rm TA}_0),\sin (\omega t-{\rm TA}_0),0\}
\]
At zero-order, the force of the Earth on the particle is
\be\label{f0}
\hspace{-2.5cm}
\mathbf{f}=\epsilon R_0 \omega^2\{f_x+e c_x \cos \omega t+ e s_x \sin \omega t, f_y+e c_y \cos \omega t+ e s_y \sin \omega t,0  \}
\ee
where $e\approx0.01671$ is the eccentricity of the Earth's orbit and 
\[
\hspace{-2cm}
\epsilon=\frac{M_\oplus}{4 M_\odot \sqrt{2-2\cos{{\rm TA}_0}}} ; \quad f_x=-2; \quad f_y=\frac{2}{\tan{{\rm TA}_0/2}};
\]
\[
\hspace{-2cm}
c_x=\frac{2}{\cos{{\rm TA}_0}-1}-1;\quad c_y=\frac{1}{\tan{{\rm TA}_0/2}};
\]
\[
\hspace{-2cm}
s_x=\frac{2}{\tan{{\rm TA}_0/2}};\quad s_y=\frac{8}{\cos{{\rm TA}_0}-1}+2.
\]
\n
The new perturbation parameter  ($\epsilon=2.6 \times 10^{-4}$ for  ${\rm TA}_0=10^\circ$) is slightly different from that introduced in \eref{pertpar} to show the explicit dependence on $TA_0$. 
{The coefficients $e c_x, e c_y, e s_x, e s_y$ are much smaller than $f_x$ and $f_y$ and therefore} we neglect the terms proportional to $e \epsilon$ in \eref{f0} and solve perturbatively the HCW equations, assuming $\mathbf{r}_g=\{0,0,0\}$ as the unperturbed motion.  
We calculate the perturbation $\mathbf{r}_1(t)$ with the assumptions  $\mathbf{r}_1(t_0)=\{0,0,0\}$ and $\dot\mathbf{r}_1(t_0)=\{0,0,0\}$ where $t_0$ is the epoch at which we put the particle at $TA_0$ degrees from the Earth. Letting $t'=t-t_0$ we have

\be\label{r1g}
\hspace{-1.5cm}
\eqalign{
\mathbf{r}_1(t')  =\epsilon R_0 \{&f_x (1 - \cos \omega t' )  + 2 f_y (\omega t'-  \sin \omega t'), \cr
  & 2f_x (\sin \omega t' - \omega t') + f_y  (4 -  4 \cos\omega t' - 3/2 \,\omega^2 t'^2 ),\cr
 & 0 \}
}
\ee

\n
We transform \eref{r1g} in the inertial coordinates using \eref{inertial} and we calculate the distance $d(t)$ of the particle from the Earth using the expression \eref{epicyclic}. Finally, we expand in Taylor series the distance to the first order in $e$ and $\epsilon$

\be\label{distappr}
d(t)=d_0+e\, d_1(t)+\epsilon\, d_2(t) + O(e^2)
\ee
where

\begin{equation*}
\hspace{-2cm}
\eqalign{
d_0=R_0 \sqrt{ 2-2\cos {\rm TA}_0}\cr
d_1(t)=\frac{R_0}{ \sqrt{ 2-2\cos {\rm TA}_0}} \left[(\cos {\rm TA}_0-1) \cos \omega t+2\sin {\rm TA}_0 \sin \omega t \right]\cr
d_2(t')=\frac{R_0}{2\sqrt{2-2\cos {\rm TA}_0}} \times
}
\end{equation*}

\begin{equation*}
\hspace{-0.5cm}
\eqalign{
 \times  \quad f_x + 2 f_y \omega t' - f_x \cos {\rm TA}_0 - 2 f_y \omega t' \cos {\rm TA}_0 +1/2 (-8 f_y + 4 f_x \omega t' +\cr
+ 3 f_y \omega^2 t'^2) \sin {\rm TA}_0 + \cos \omega t'  (-f_x + f_x \cos {\rm TA}_0 + 4 f_y \sin {\rm TA}_0) +\cr
 + (-2 f_y + 2 f_y \cos {\rm TA}_0 - 2 f_x \sin {\rm TA}_0) \sin \omega t' .
}
\end{equation*}

\n
The term $d_0$ is a constant, the term $d_1$ is a sum of sinusoids with 1year period. The $d_2$ term contains linear and quadratic terms in $t-t_0$,  and is therefore negligible for $t\approx t_0$ because $\epsilon \ll e$ but it becomes dominant for larger $t$.
\n
The epochs  of the relative minima and maxima of $d(t)$ depend on TA$_0$ but not on $t_0$.
They are found by equating to zero the first derivative of $d_1$:

\begin{equation*}
\eqalign{
2 \cos \omega t \sin {\rm TA}_0 +(1-\cos {\rm TA}_0) \sin \omega t =0.\cr
}
\end{equation*}
With the additional condition on the second derivative
\begin{equation*}
(1-\cos {\rm TA}_0)\cos \omega t_{min}-2 \sin {\rm TA}_0 \sin \omega t_{min} >0,
\end{equation*}
the minima occur at
\be\label{t0}
t_{min,k}=-\frac{1}{\omega}\arctan \left[\frac{2}{\tan({\rm TA}_0/2)}\right]+\frac{2 k \pi}{\omega}, \qquad  k \in \mathbb{Z}.
\ee
\n
In the same fashion of \eref{distappr}, the TA can be obtained, to first order in $e$, as 

\be
\hspace{-2.5cm}
{\rm TA}(t)={\rm TA}_0+2 e \sin \omega t+\epsilon \left[4 f_y(\cos \omega t' -1)+2 f_x (\omega t'-\sin \omega t')+\frac{3}{2}f_y (\omega t')^2 \right].
\label{last}
\ee
%and the minima occur at slightly different epochs
%\[
%t_{min}=\frac{3}{2 \omega} \pi+\frac{2 k \pi}{\omega}.
%\]
\n
Although the epochs $t_{min}$ are obtained using a first-order approximation, they are in good agreement with the exact values (for an example, see the bottom panels of \Fref{OPTJPLIRT}  and  \ref{OPTJPLET}).

\section*{Acknowledgments}
We  thank Oliver Jennrich, Pete Bender and Bill Weber for useful discussions.

\end{document}